\documentclass[11pt]{article}
\usepackage{latexsym}  
\usepackage{amssymb}
\usepackage{graphicx}
\usepackage{amsmath}

\begin{document}

\begin{center}
{\Large\bf Phenomenological Meaning of a Neutrino Mass Matrix }\\[.1in]
{\Large\bf  Related to Up-Quark Masses}

\vspace{3mm}
{\bf Yoshio Koide}

{\it IHERP, Osaka University, 1-16 Machikaneyama, 
Toyonaka, Osaka 560-0043, Japan} \\
{\it E-mail address: koide@het.phys.sci.osaka-u.ac.jp}

\date{\today}
\end{center}

\vspace{3mm}
\begin{abstract}
Recently, a curious neutrino mass matrix has been proposed: 
it is related to up-quark masses,
and it can excellently give a nearly tribimaxial mixing. 
It is pointed out that, in order to obtain such successful 
results, three phenomenological relations among masses and 
CKM parameters must be simultaneously satisfied.
This suggests that there must be a specific flavor-basis
in which down-quark and charged lepton mass matrices are
simultaneously diagonalized.
\end{abstract}

\vspace{3mm}

{\large\bf 1 \ Introduction}
 
Recently, a curious neutrino mass matrix has been proposed by the 
author \cite{Koide-O3}: 
the mass matrix is related to up-quark masses as follows:
$$
M_\nu = M_D M_R^{-1} M_D^T ,
\eqno(1.1)
$$
where the neutrino Dirac mass matrix $M_D$ is given by 
$M_D \propto M_e$  
($M_e$ is a charged lepton mass matrix),
and the right-handed neutrino Majorana mass matrix $M_R$ is given by
$$
M_R \propto M_e M_u^{1/2} + M_u^{1/2} M_e .
\eqno(1.2)
$$
The mass matrix (1.1) with (1.2) has been derived from
an idea that the origin of the mass spectra (i.e.  
effective Yukawa coupling constants) 
is due to vacuum expectation values (VEV) structures of 
gauge singlet scalars $\Phi_{ij}$.
(The details are reviewed in the next section.)
In order to obtain the lepton mixing matrix $U$, one
must know forms of $M_D$ and $M_u^{1/2}$ 
in the ``$e$-basis" (we refer to a diagonal basis of the
mass matrix $M_f$  as ``$f$-basis"). 
The form $M_D=M_e$ is given by
$M_e ={\rm diag}(m_e, m_\mu, m_\tau)$ in the $e$-basis.
For the form $M_u^{1/2}$, by analogy with the relation
$M_u=V^T D_u V$ in the $d$-basis, where 
$D_u ={\rm diag}(m_u, m_c,m_t)$ and $V$ is the 
Cabibbo-Kobayashi-Maskawa (CKM) quark mixing matrix
(and note that a mass matrix $M_f$ is diagonalized as
a form $U_f^T M_f U_f = D_f$ in the present model
because we assume an O(3) flavor symmetry as we 
mention it in Sec.2),
we assume that $M_u^{1/2}$ in the $e$-basis is given by
a form
$$
M_u^{1/2} = V^T(\delta) D_u^{1/2} V(\delta) ,
\eqno(1.3)
$$
where
we have adopted the standard expression $V{SD}$ \cite{sdCKM} 
as a phase convention of the CKM matrix $V(\delta)$.
In order to estimate the form $M_\nu$, we use the following 
observed up-quark masse values at an energy scale of the
weak interactions $\mu=m_Z$ 
\cite{Xing-qmass}, $m_u=0.00127$ GeV, $m_c=0.619$ GeV, 
$m_t=171.7$ GeV, and the observed CKM mixing parameters 
(best-fit values) \cite{PDG08} $|V_{us}|=0.2257$, 
$|V_{cb}|=0.0415$ and $|V_{ub}|=0.00359$ 
together with the observed charged lepton masses. 
(Here, since we use the values at $\mu=m_Z$ for the CKM matrix 
parameters, we also use the running mass values at $\mu=m_Z$.)
Then, one can successfully obtain
a nearly tribimaximal mixing \cite{tribi}, 
$$
U= \left(
\begin{array}{ccc}
+0.8026 & -0.5966 & -0.0009 \\
-0.4356 & -0.5871 & +0.6823 \\
+0.4076 & +0.5472 & +0.7311
\end{array} \right) ,
\eqno(1.4)
$$
for $\delta=\pi$, i.e.
$$
\sin^2 2\theta_{23}=0.9952, \ \ \ \tan^2 \theta=0.5525, \ \ \ 
|U_{13}|= 0.00094 .
\eqno(1.5)
$$
For reference, we give phase-dependence of the numerical 
results in Table 1.
The best-fit values \cite{PDG08} of the CKM mixing parameters 
show $\delta=69.8^\circ$. 
However, as seen in Table 1, the predicted value of 
$\sin^2 2\theta_{23}$ at  $\delta \simeq 69.8^\circ$
is in poor agreement with the observed value 
 $\sin^2 2\theta_{23}=1.00_{-0.13}$ \cite{MINOS},
 although the predicted value of $\tan^2\theta_{12}$ is
 roughly in agreement with the observed value 
$\tan^2\theta_{12}=0.47^{+0.06}_{-0.05}$ \cite{KamLAND}. 
As stated in the next section, since the flavor-basis transformation
matrix is confined to an orthogonal matrix because the present model
is based on an O(3) flavor symmetry, the phase parameter
$\delta$ must be $0$ or $\pi$.

We also list numerical results for the original Kobayashi-Maskawa 
phase convention \cite{KM} in Table 2.
As seen in Table 2, not only the both cases, $\delta=0$ and 
$\delta=\pi$, but also any values of $\delta$ 
cannot give a reasonable value of $\sin^2 2\theta_{23}$.
Thus, we find that the phenomenological success is only for
the case of $V(\delta)=V_{SD}(\delta)$ (not for the original KM phase 
convention of CKM matrix).

\begin{table}
\caption{
$\delta$ dependence of predicted values in the standard
phase convention of $V(\delta)$.
Here, $|V_{us}|$, $|V_{cb}|$ and $|V_{ub}|$ 
have been used as three input values of the four
independent parameters of $V(\delta)$.
The  best-fit value of $\delta$ in the quark sector is 
$\delta_q=69.8^\circ$ 
from the observed CKM matrix data. 
}

\begin{center}
\begin{tabular}{|c|l|l|l|l|l|}\hline
$\delta$ &  $\sin^2 2\theta_{23}$ & $\tan^2 \theta_{12}$ &
$|U_{13}|$ & $\Delta m^2_{21}/\Delta m^2_{32}$ \\ \hline
0    &  $0.4803$ & $0.4745$ & $0.01042$ & $0.00196$ \\
$60^\circ$ & $0.7631$ & $0.4801$ & $0.00844$ & $0.00139$ \\
$69.8^\circ$ & $0.8127$ & $0.4851$ & $0.00781$ & $0.00127$ \\
$90^\circ$ & $0.9028$ & $0.5017$ & $0.00615$ & $0.00102$ \\
$120^\circ$ & $0.9688$ & $0.5277$ & $0.00386$ & $0.00081$ \\
$180^\circ$ &  $0.9952$ & $0.5525$ & $0.00094$ & $0.00068$ 
\\ \hline
\end{tabular}\end{center} 
\end{table}

\begin{table}
\caption{
$\delta$ dependence of predicted values in the original
Kobayashi-Maskawa phase convention of $V(\delta)$.
Here, $|V_{us}|$, $|V_{td}|$ and $|V_{ub}|$ 
have been used as three input values of the four
independent parameters of $V(\delta)$.
The best-fit value of $\delta$ in the quark sector
is $\delta_q=90.8^\circ$
from the observed CKM matrix data. 
}

\begin{center}
\begin{tabular}{|c|l|l|l|l|l|}\hline
$\delta$ &  $\sin^2 2\theta_{23}$ & $\tan^2 \theta_{12}$ &
$|U_{13}|$ & $\Delta m^2_{21}/\Delta m^2_{32}$ \\ \hline
0    &  $0.7821$ & $0.5074$ & $0.00769$ & $0.00093$ \\
$60^\circ$ & $0.8088$ & $0.3587$ & $0.0303$ & $0.0052$ \\
$90^\circ$ & $0.8781$ & $0.1862$ & $0.0614$ & $0.04269$ \\
$120^\circ$ & $0.8482$ & $0.3523$ & $0.03303$ & $0.00752$ \\
$180^\circ$ &  $0.8369$ & $0.5028$ & $0.00329$ & $0.00169$ 
\\ \hline
\end{tabular}\end{center} 
\end{table}

In order to obtain the phenomenological success,
it is essential to assume not only the neutrino mass matrix
form (1.1) with (1.2), but also forms of flavor-basis
transformation matrices $U_{ud}$ and $U_{ue}$ 
$$
\begin{array}{l}
U_{ud}=V_{SD}(\delta_q) \ \ \ (\delta_q \simeq 70^\circ) , \\
U_{ue}=V_{SD}(\delta_\ell) \ \ \ (\delta_\ell = 180^\circ) ,
\end{array}
\eqno(1.6)
$$
where $U_{ff'}$ transforms a matrix in an $f$-basis into 
that in an $f'$-basis, and 
$V_{SD}(\delta)$ is the standard phase convention of
the CKM matrix with the observed values of $|V_{us}|$,
$|V_{cb}|$ and $|V_{ub}|$ as three input values of the four
independent parameters of $V_{SD}$.

In Sec.2, we give a short review of the model which leads
to the mass matrix (1.1) with (1.2).
In Sec.3, we investigate relations between
conditions for tribimaximal mixing and the empirical
neutrino mass matrix (1.1) with (1.2) and (1.3)
from the phenomenological point of view. 
One will find that three phenomenological relations
among the masses and CKM matrix parameters must 
simultaneously be satisfied in order to get a nearly
tribimaximal mixing.
As we state in Sec.3, 
it is hard to consider that such the simultaneous 
coincidence are accidental, so that it should be
considered that such phenomenological relations 
originate in a common law.
In Sec.4, we speculate possible forms of the mass matrices
$M_d$ and $M_e$.
It is concluded that there is a flavor basis in which 
the down-quark and charged lepton mass matrices, $M_d$ and
$M_e$, are simultaneously diagonalized.
Finally, Sec.5 is devoted to summary and remarks.  


\vspace{3mm}

{\large\bf 2 \ Model}

The neutrino mass matrix related to up-quark masses has first 
been derived on the basis of a U(3) flavor symmetry model 
\cite{empiricalMnu08}, 
and then, the form (1.1) with (1.2) has
been derived on the basis of an O(3) flavor symmetry model 
\cite{Koide-O3}.
In this section, we give a short review of the O(3) model.
 
It is assumed that effective Yukawa coupling constants
$Y_f^{eff}$ are given by VEVs $\langle Y_f\rangle$ 
of gauge singlet scalars $Y_f$ (for convenience, we refer to
those fields as ``Yukawaons")
which belong to $({\bf 3}\times{\bf 3})_S ={\bf 1}+{\bf 5}$ 
of an O(3) flavor symmetry:
$$
W_{Y}= \sum_{i,j} \frac{y_u}{\Lambda} U_i(Y_u)_{ij} {Q}_{j} H_u  
+\sum_{i,j}\frac{y_d}{\Lambda} D_i(Y_d)_{ij} {Q}_{j} H_d 
$$
$$
+\sum_{i,j} \frac{y_\nu}{\Lambda} L_i(Y_\nu)_{ij} {N}_{j} H_u  
+\sum_{i,j}\frac{y_e}{\Lambda} L_i(Y_e)_{ij} E_j H_d  
+h.c. + \sum_{i,j}y_R N_i (Y_R)_{ij} N_j ,
\eqno(2.1)
$$ 
where $Q$ and $L$ are quark and lepton 
SU(2)$_L$ doublet fields of O(3)$_F$ triplets,  
and $U$, $D$, $N$, and $E$ are 
SU(2)$_L$ singlet matter fields of O(3)$_F$ triplets,
and $\Lambda$ is an energy scale of an effective theory.
Since we assume the O(3) flavor symmetry, 
the Yukawaons $Y_f$ ($f=u,d,e,\nu,R$) are symmetric.
Under this definition of $(Y_f)_{ij}$ given by Eq.(1.1),
the VEV matrix $\langle Y_f \rangle$ are diagonalized
as $U_f^T \langle Y_f \rangle U_f =\langle Y_f \rangle^D$,
where the index $D$ means that the matrix is diagonal,
and the quark and lepton mixing matrices $V$ and $U$ are
given by $V=U_u^\dagger U_d$ and $U =U_e^\dagger U_\nu$,
respectively.
In order to distinguish the Yukawaons $Y_f$ from each other, 
the following U(1)$_X$ charges are assigned:  $Q_X(Y_f)=x_f$ 
($f=u,d,\nu,e$), $Q_X(U)=-x_u$, $Q_X(E)=-x_e$, and so on.
The field $Y_R$ has a charge $Q_X(Y_R)=2x_\nu$.

One writes a superpotential $W$ under the following conditions:
(i) Terms consist of, at most, holomorphic cubic terms of the 
fields, and do not contain higher dimensional terms, except for 
the Yukawa interaction terms $W_Y$; 
(ii) Those are invariant under the O(3)$_F$ and U(1)$_X$ symmetries.
(iii) Yukawaons $Y_f$ always behave as a combination of {\bf 1}+{\bf 5},
so that, for example, {\bf 5} alone never appears
in the interaction terms.

The VEV spectra $\langle Y_f \rangle$
are evaluated from supersymmetric (SUSY) vacuum conditions
for a superpotential $W=W_Y+W_u+W_d+W_e+W_\nu+W_R$,
where $W_f$ ($f=u,d,\nu,e,R$) play a role in fixing the VEV 
structures $\langle Y_f\rangle$.
(Since one can easily show $\langle Q\rangle =\langle L\rangle 
=\langle U\rangle =\langle D\rangle =\langle N\rangle 
=\langle E\rangle =0$, hereafter, the term $W_Y$ is dropped from 
the superpotential $W$ as far as the VEV structures of $Y_f$
are investigated.)
For example, a spectrum of $\langle Y_u\rangle$ is obtained
from the following superpotential terms $W_u$:
$$
W_u = \lambda_u {\rm Tr}[\Phi_u \Phi_u A_{u}]
+\mu_u {\rm Tr}[Y_u A_{u}] + W_{\Phi u} ,
\eqno(2.2)
$$
where a new filed $A_u$ has U(1)$_X$ charge $Q_X=-x_u$.
Here, the term $W_{\Phi u}$ has been introduced in order to
fix eigenvalues of $\langle \Phi_u \rangle$.
Since the purpose of the present paper is not to discuss
quark and lepton mass spectra, an explicit form of $W_{\Phi u}$ 
is given in Appendix A.
Since $W_{\Phi u}$ contains $Y_u$ and $\Phi_u$ as shown in
Appendix A, SUSY vacuum conditions $\partial W/\partial Y_u=0$ 
and $\partial W/\partial \Phi_u=0$
will be discussed in Appendix A. 
From a SUSY vacuum condition $\partial W/\partial A_u=0$ 
(for the moment, one regards $W_u$ as $W$), one obtains
$$
\frac{\partial W}{\partial A_{u}} = 0
=\lambda_u \Phi_u \Phi_u + \mu_u Y_u ,
\eqno(2.3)
$$
so that one obtains a bilinear relation
$$
\langle Y_u\rangle = - \frac{\lambda_u}{\mu_u} 
\langle\Phi_u\rangle \langle\Phi_u\rangle ,
\eqno(2.4)
$$
i.e. the field $\Phi_u$ plays a role of $M_u^{1/2}$
in Eq.(1.2).
For convenience, we refer to $\Phi_f$ as ``ur-Yukawaons". 
The ur-Yukawaons $\Phi_f$ play a role in fixing VEV spectra of
Yukawaons.
Although we consider 5 Yukawaons $Y_f$ ($f=u, d, e, \nu, R$), 
we will consider only 2 ur-Yukawaons $\Phi_e$ and $\Phi_u$
in the present model.
Note that, since the matrix $\langle \Phi_u \rangle$ is
not Hermitian, the relation
$$
U_u^T \langle Y_u \rangle U_u =
\langle Y_u \rangle^D \propto {\rm diag} ({m_{u}},
{m_{c}},{m_{t}}) ,
\eqno(2.5)
$$
does not always mean
$$
U_u^T \langle \Phi_u \rangle U_u =
\langle \Phi_u \rangle^D \propto {\rm diag} (\sqrt{m_{u}},
 \sqrt{m_{c}},\sqrt{m_{t}}) ,
\eqno(2.6)
$$ 
where $D$ denotes that the matrix is  diagonal.
As one sees later, one needs the relation (2.6).
Therefore, we assume the field $\Phi_u$ (and also $Y_u$)
is real, so that the matrix $U_u$ is orthogonal matrix.

For the charged lepton sector, we also assume 
superpotential term $W_e$ similar to the up-quark 
sector:
$$
W_e = \lambda_e {\rm Tr}[\Phi_e \Phi_e A_{e}]
+\mu_e {\rm Tr}[Y_e A_{e}] + W_{\Phi e} ,
\eqno(2.7)
$$
where $\Phi_e$, $Y_{e}$ and $A_e$ have U(1)$_X$ charges
$\frac{1}{2}x_e$, $x_e$ and $-x_e$, respectively, 
so that one obtains relations
$$
Y_e = - \frac{\lambda_e}{\mu_e} \Phi_e\Phi_e ,
\eqno(2.8)
$$
with $\Phi_e^D \propto {\rm diag} (\sqrt{m_{e}},
 \sqrt{m_{\mu}},\sqrt{m_{\tau}})$,
where one has again assumed that the field $\Phi_e$ is real.
(Here and hereafter, for simplicity,  we will sometimes express 
VEV matrices $\langle M\rangle$ as simply $M$.)

Next, let us  investigate a possible form of $W_R$.
We introduce a new field $A_R$ with U(1)$_X$ charge 
$Q_X= -2x_\nu$. 
In order to obtain the relation (1.2), we assume the following form 
of $W_R$:
$$
W_R = \lambda_R  {\rm Tr}[(Y_e \Phi_u +\Phi_u Y_e) A_R] 
+\mu_R {\rm Tr}[Y_R A_R] + \lambda_R \xi {\rm Tr}[Y_\nu Y_\nu A_R] ,
\eqno(2.9)
$$
where we have assumed a relation among the U(1)$_X$ charges,
$$
2x_\nu= \frac{1}{2}x_u +x_e  .
\eqno(2.10)
$$  
From SUSY vacuum conditions 
${\partial W}/{\partial {Y}_{R} } = 0$,  one obtains $A_R=0$.
Then, the requirement ${\partial W}/{\partial Y_e} = 0$ 
leads to the condition
${\partial W_e}/{\partial Y_e}=0$, so that one obtains
the relation (2.8).
From ${\partial W}/{\partial {A}_R } = 0$, one obtains
$$
Y_R = - \frac{\lambda_R}{\mu_R} ( Y_e \Phi_u + \Phi_u Y_e
+\xi Y_\nu Y_\nu) .
\eqno(2.11)
$$
The third term ($\xi$-term) does not affect a form of
the lepton mixing matrix $U$ because the term gives
a constant term proportional to a unit matrix ${\bf 1}$
as shown later. 
Thus, one can obtain the desirable form (1.2) of $Y_R$.

Next, we discuss how to obtain $\langle Y_\nu\rangle
= \langle Y_e\rangle$.
The simplest assumption to obtain a relation 
$M_D \propto M_e$ (i.e. $Y_\nu \propto Y_e$) is to assume 
that the fields $N$ and $E$ have the same U(1)$_X$ charges
(i.e. $x_\nu=x_e$), and to 
consider a model without $Y_\nu$. 
However, then, one obtains $x_\nu =x_e = x_u/2$ from the
relation (2.10), so that $Y_e$ and $\Phi_u$ (and also
$Y_u$ and $Y_R$) have the same U(1)$_X$ charges.
This brings some additional terms into $W_u$, $W_e$ and $W_R$ 
due to the mixings between $Y_e$ and $\Phi_u$ and between 
$Y_u$ and $Y_R$, so that one cannot obtain 
desirable relations without ad hoc selections of those terms. 
Therefore, in order to obtain the relation 
$Y_\nu \propto Y_e$ with $x_\nu \neq x_e$,
we assume the following structure of $W_\nu$:
$$
W_\nu = \lambda_{\nu\nu} \phi_\nu {\rm Tr}[Y_\nu A_\nu]
+\lambda_{\nu e} \phi_e {\rm Tr}[Y_e A_\nu],
\eqno(2.12)
$$
where $\phi_\nu$ and $\phi_e$ are gauge- and
flavor-singlet fields, and  we assign U(1)$_X$ charges
as $Q_X(A_\nu)=x_\nu$, $Q_X(\phi_\nu)=-(x_\nu+x_\nu)$ and
$Q_X(\phi_e)=-(x_e+x_\nu)$.
From $\partial W/\partial \phi_\nu=0$ and 
$\partial W/\partial \phi_e=0$, one obtains 
$A_\nu =0$.
From $\partial W/\partial A_\nu=0$, one obtains
$$
Y_\nu = -\frac{\lambda_{\nu e} \phi_e}{\lambda_{\nu\nu} \phi_\nu}
Y_e .
\eqno(2.13)
$$

In order to obtain a neutrino mixing matrix form 
in the $e$-basis, 
one must know a matrix form of $\langle\Phi_u\rangle$ in 
the $e$-basis, 
although the form $\langle\Phi_u\rangle^D$ on the $u$-basis 
is given by Eq.(2.6).
(Now, ``$f$-basis" is defined as a flavor basis in which
the VEV matrix $\langle Y_f\rangle$ is diagonal.)
Let us define a transformation of a VEV matrix 
$\langle Y_f\rangle$ 
from  a $b$-basis to an $a$-basis as
$$
\langle Y_f\rangle_a = U_{ba}^T \langle Y_f\rangle_b U_{ba} ,
\eqno(2.14) 
$$
where $U_{ab}$ are unitary matrices, and $\langle Y_f\rangle_a$
denotes a VEV matrix form on the $a$-basis. 
The unitary matrices $U_{ab}$ satisfy
$U_{ab}^\dagger =U_{ba}$ and $U_{ab}U_{bc}=U_{ac}$.
(These operators $U_{ab}$ are not always members of
O(3) flavor-basis transformations.)  
Since $Y_f^T=Y_f$ in the present model, the VEV matrix 
$\langle Y_f\rangle$ are diagonalized as
$U_f^T \langle Y_f\rangle U_f = \langle Y_f\rangle^D
\equiv \langle Y_f\rangle_f$.
Therefore, $\langle Y_u\rangle_d$ is given by
$$
\langle Y_u\rangle_d =V^T(\delta) \langle Y_u\rangle_u
V(\delta) ,
\eqno(2.15)
$$
where $V(\delta)$ is the standard expression of the 
CKM matrix.
The simplest assumption is to consider that the $d$-basis 
is identical with the $e$-basis, and then,  
one can regard $U_{ue}$ as $U_{ue} =V$ because $U_{ud}= V$.
However, since one has assumed that $Y_u$ and $Y_e$ are
real, the flavor-basis transformation matrix $U_{ue}$
must be orthogonal, i.e. the phase parameter $\delta$
is $0$ or $\pi$ even if one assumes the form 
$U_{ue}= V(\delta)$.
As one has already seen in Table 1, the case with 
$\delta=\pi$ can give reasonable numerical results.

Anyhow, one assumes the form 
$$
\langle\Phi_u\rangle_e =U_{ue}^T \langle\Phi_u\rangle_u U_{ue} =
V^T(\delta) \langle\Phi_u\rangle^D V(\delta) ,
\eqno(2.16)
$$
one can obtain the following phenomenological neutrino mass 
matrix
$$
\langle M_\nu \rangle_e = k_\nu Y_e^D \left[ 
 V^T(\delta)\Phi_u^D V(\delta) Y_e^D
+ Y_e^D  V^T(\delta)\Phi_u^D V(\delta) 
+ \xi Y_\nu Y_\nu \right]^{-1} Y_e^D 
$$
$$
=k_\nu \left[ (Y_e^D)^{-1} 
 V^T(\delta)\Phi_u^D V(\delta) 
+   V^T(\delta)\Phi_u^D V(\delta) (Y_e^D)^{-1}
+ \xi_0 {\bf 1} \right]^{-1}  ,
\eqno(2.17)
$$
where 
$Y_e^D \propto {\rm diag}(m_e, m_\mu, m_\tau)$ and
$\Phi_u^D \propto {\rm diag}(\sqrt{m_u}, \sqrt{m_c}, 
\sqrt{m_t})$.
The third term ($\xi_0$ term) does not affect the 
lepton mixing matrix $U$.
Rather, the existence of the $\xi_0$ term is useful
to adjust the value of $\Delta m^2_{21}/\Delta m^2_{32}$
because the predicted values in Table 1 were 
considerably small compared to the observed value
$|R|=0.028\pm 0.004$,
where one has used the observed values 
$\Delta m^2_{21}=(7.59\pm 0.21) \times 10^{-5}\, {\rm eV}^2$
\cite{KamLAND} and 
$|\Delta m^2_{32}|=(2.74^{+0.44}_{-0.26}) \times 10^{-3}\, {\rm eV}^2$
\cite{MINOS}.

\vspace{3mm}

{\large\bf 3 \ Conditions for a tribimaximal mixing}

In this section, we investigate what phenomenological relations 
are required for the mass matrix (2.17) in order to give 
a nearly tribimaximal mixing. 
Since one know \cite{inverse} that a mixing matrix for 
$(M_\nu)^{-1}$ is given by $U^*$
when a mixing matrix for $M_\nu$ is given by $U$, 
for the purpose to obtain conditions for a tribimaximal
mixing, one can investigate the following matrix 
$$
M= (Y_e^D)^{-1} 
 V^T(\delta)\Phi_u^D V(\delta) 
+   V^T(\delta)\Phi_u^D V(\delta) (Y_e^D)^{-1}
+ \xi_0 {\bf 1} ,
\eqno(3.1)
$$
i.e.
$$
M_{ij} = \left( \frac{1}{m_{e i}} + \frac{1}{m_{e j}} \right)
\sum_k \sqrt{m_{u k}} V_{ki} V_{kj} ,
\eqno(3.2)$$
instead of the mass matrix (2.17).
Since the $\xi_0$-term is not essential for evaluating
the mixing matrix $U$, hereafter, we put $\xi_0=0$.
(Although a similar study has been done in Ref.\cite{empiricalMnu08}
based on a U(3) flavor symmetry, where the VEV matrix 
$\langle\Phi_u\rangle_e$ has been given by 
$\langle\Phi_u\rangle_e=V^\dagger(\delta) 
\langle\Phi_u\rangle_u V(\delta)$, in the present O(3) model,
the VEV matrix $\langle\Phi_u\rangle_e$ is given by 
$\langle\Phi_u\rangle_e=
V^T(\delta) \langle\Phi_u\rangle_u V(\delta)$.)

As shown in Appendix, the conditions to obtain the maximal
$2\leftrightarrow 3$ mixing, i.e.
$$
\sin^2 2\theta_{23} \equiv 4|U_{23}|^2 |U_{33}|^2 =1, \ \ 
\ |U_{13}|^2=0 ,
\eqno(3.3)
$$
are
$$
|M_{12}| = |M_{13}|,
\eqno(3.4)
$$
and
$$
|M_{22}| = |M_{33}|,
\eqno(3.5)
$$
From Eq.(3.2), one obtains
$$
M_{12} \simeq \frac{\sqrt{m_{c}}}{m_{e}} 
 V_{21}V_{22} ,
\eqno(3.6)
$$
$$
M_{13} \simeq \frac{\sqrt{m_{t}}}{m_{e}} V_{31}V_{33} ,
\eqno(3.7)
$$
$$
M_{22} \simeq 2 \frac{\sqrt{m_{t}}}{m_{\mu}} V_{33}^2 ,
\eqno(3.8)
$$
$$
M_{33} \simeq 2 \frac{\sqrt{m_{t}}}{m_{\tau}} V_{33}^2 ,
\eqno(3.8)
$$
where one has assumed a hierarchical structure of $|V_{ij}|$
similar to the observed CKM matrix.
The condition (3.5) requires 
$$
\sqrt{ \frac{m_c}{m_t} } \simeq \frac{m_\mu}{m_\tau} ,
\eqno(3.9)
$$
The left- and right-hand sides of Eq.(3.9) give  values
\cite{Xing-qmass} $0.060$ and $0.059$, respectively.  
Therefore, the condition
(3.5) is phenomenologically well satisfied.
On the other hand, the condition (3.4) requires 
$$
\sqrt{ \frac{m_c}{m_t} } \simeq \frac{|V_{31}|}{|V_{21}|} .
\eqno(3.10)
$$
In order to evaluate the relation (3.10),
one uses a relation
$$
\frac{V_{31}}{V_{21}}= -\left( \frac{V_{23}^*}{V_{33}^*}
+\frac{V_{11}}{V_{21}} \frac{V_{13}^*}{V_{33}^*} \right) ,
\eqno(3.11)
$$
from the unitary relation $V_{11}V_{13}^* +V_{21}V_{23}^* +
V_{31} V_{33}^*=0$.
For a standard expression of the CKM matrix, Eq.(3.11)
leads to
$$
\frac{V_{31}}{V_{21}} \simeq -\left( |V_{cb}|-
\frac{|V_{ub}|}{|V_{us}|} e^{i\delta} \right) .
\eqno(3.12)
$$
The left-hand side of Eq.(3.10) is $0.060$, and the
right-hand side is $0.0412+0.0174=0.0586$ for 
$\delta=\pi$. 
Thus, the condition is also well satisfied.
Note that if the observed value of $|V_{td}|$, 
$|V_{td}|=0.00874$ \cite{PDG08}, as the
value $|V_{31}|$ is used, 
the condition (3.10) cannot be satisfied.
This is a reason for that 
when one used the original Kobayashi-Maskawa phase 
convention instead of the standard CKM matrix expression, 
one could not give a nearly tribimaximal mixing 
as seen in Table 2. 

Next, we check the condition to give $\tan^2 \theta_{12}=1/2$,
$$
\eta^2 \left( (M_{22} M_{33} )^{1/2} +M_{23} \right) -M_{11}
=\eta ({M}_{12} M_{13})^{1/2} ,
\eqno(3.13)
$$   
(see (B.16) in Appendix B).
From Eqs.(3.6) - (3.8) and 
$$
M_{11} \simeq 2 \frac{ \sqrt{m_{t}}}{m_{e}} 
\left( \sqrt{ \frac{m_{c}}{m_{t}}} V_{21}^2 
+ \sqrt{ \frac{m_{u}}{m_{t}}} V_{11}^2 
\right) ,
\eqno(3.14)
$$
$$
M_{23} \simeq  \frac{ \sqrt{m_{t}}}{m_{\mu}} V_{32} V_{33} ,
\eqno(3.15)
$$
one finds $|M_{22}+M_{23}| \ll |M_{11}$, so that the condition
(3.13) requires $M_{23} \simeq M_{11}$ ($\eta=-1$ in the present
case).
The condition $M_{23}\simeq M_{11}$ requires
$$
|V_{us}|+\frac{1}{|V_{us}|}
\sqrt{ \frac{m_{u}}{m_{t}}} \simeq \frac{1}{2} .
\eqno(3.16)
$$
The left-had side of Eq.(3.16) gives $0.2257+02007=0.4264$.
Considering the present rough approximation, one may
consider that the condition (3.13) is roughly satisfied.

In conclusion, in order to obtain a tribimaximal mixing,
the three phenomenological relations (3.9), (3.10) and
(3.16) must simultaneously be satisfied.
It is hard to consider that such the simultaneous 
coincidences are accidental.
Rather, it should be considered that such the 
phenomenological relations originate in a common law.
Also, one must note that, in order to satisfy the condition 
(3.10), one must take the standard expression of the CKM
matrix and use the observed values $|V_{us}|$, $|V_{cb}|$
and $|V_{ub}|$ in order to fix the three rotation angles
in the CKM matrix.
This suggests that the down-quark mass
matrix $M_d$ has a similar structure with the charged
lepton mass matrix $M_e$.
In the next section, we will investigate  
a possible relation between $M_d$ and $M_e$.

\vspace{3mm}

{\large\bf 4 \ Possible structures of $M_d$ and $M_e$}

In this section, we speculate possible mass matrix forms of 
the down-quark and charged lepton mass matrices $M_d$ and 
$M_e$ which lead to the assumption (1.6).

Generally, there are 9 phase conventions of the CKM matrix $V$
\cite{9CKM}:
$$
V(m,n) =R_m P_\ell R_\ell R_n \ \ \ \ \ (m\neq \ell \neq n),
\eqno(4.1)
$$
where $m,n,\ell = 1,2,3$, and
$$
R_1 (\theta) = \left(
\begin{array}{ccc}
1 & 0 & 0 \\
0 & c & s \\
0 & -s & c 
\end{array} \right) , \ \ \ \ 
R_2 (\theta) = \left(
\begin{array}{ccc}
c & 0 & s \\
0 & 1 & 0 \\
-s & 0 & c 
\end{array} \right) , \ \ \ \ 
R_3 (\theta) = \left(
\begin{array}{ccc}
c & s & 0 \\
-s & c & 0 \\
0 & 0 & 1 
\end{array} \right) ,
\eqno(4.2)
$$
$$
\begin{array}{l}
P_1 = {\rm diag} (e^{i \delta}, \ 1, \ 1) , \\ 
P_2 = {\rm diag} (1, \ e^{i \delta}, \ 1), \\ 
P_3 = {\rm diag} (1, \ 1, \ e^{i \delta}) .
\end{array}
\eqno(4.3)
$$
($c=\cos\theta$ and $s=\sin\theta$).
For example, the standard expression $V_{SD}$ of the CKM matrix
$$
V_{SD}(\delta) = R_1(\theta_{23}) P_3(\delta) R_2(\theta_{13})
P_3(-\delta) R_3(\theta_{12}) ,
\eqno(4.4)
$$
is rewritten as
$$
V_{SD}(\delta) =e^{i\delta} P_1(-\delta) R_1(\theta_{23}) 
P_2(-\delta) R_2(\theta_{13})
 R_3(\theta_{12})P_3(-\delta) ,
\eqno(4.5)
$$
because $P_3(\delta)=e^{i\delta} P_1(-\delta) P_2(-\delta)$.
Since the factors $e^{i\delta} P_1(-\delta)$ and $P_3(-\delta)$
in the left- and right-hand sides can be absorbed into the
unobservable phases of up- and down-quarks, respectively, 
the standard expression $V_{SD}$ corresponds to the 
expression $V(1,3)$ defined in (4.1).

In the O(3) model, where the mass matrices are symmetric, 
the mass matrices $M_u$ and $M_d$ are diagonalized as
$$
U_u^T M_u U_u = D_u , \ \ \   U_d^T M_d U_d = D_d ,
\eqno(4.6)
$$
and the CKM matrix $V$ is given by
$$
V=U_u^\dagger U_d .
\eqno(4.7)
$$
As seen in the general expressions of $V$ given in (4.1),
one can always find a flavor basis (we refer to it as a 
``$x$-basis") in which the $CP$-violating phases are 
factorized as
$$
\langle Y_u \rangle_x = P_n(\delta_u) 
\langle \widetilde{Y}_u \rangle_x P_n(\delta_u) , \ \ \ 
\langle Y_d \rangle_x = P_n(\delta_d) 
\langle \widetilde{Y}_d \rangle_x P_n(\delta_d) ,
\eqno(4.8)
$$
where $\langle \widetilde{Y}_u \rangle_x$ and 
$\langle \widetilde{Y}_d \rangle_x$ are real matrices,
and they are diagonalized by rotation matrices $R_u$ and 
$R_d$ as
$$
R_u^T \langle \widetilde{Y}_u \rangle_x R_u = D_u , \ \ \ 
R_d^T \langle \widetilde{Y}_d \rangle_x R_d = D_d , 
\eqno(4.9)
$$
respectively.
Then, since $U_u=P_n(-\delta_u)R_u$ and $U_d=P_n(-\delta_d)R_d$,
one obtains the expression of the flavor-basis transformation
operator $U_{ud}$
$$
U_{ud} =V = R_u^T P_n(\delta_u-\delta_d) R_d .
\eqno(4.10)
$$
Similarly, one can obtain an expression of $U_{ue}$ as follows:
$$
U_{ue} = R_u^T P_n(\delta_u-\delta_e) R_e .
\eqno(4.11)
$$

Now, let us return to our model.
As seen in Sec.3, the requirement (1.6) for a nearly tribimaximal 
mixing is rewritten as
$$
\begin{array}{l}
U_{ud} = R_1(\theta_{23}) P_3(\delta_q) R_2(\theta_{13})
P_3(-\delta_1) R_3(\theta_{12}) , \\
U_{ue} = R_1(\theta_{23}) P_3(\delta_\ell) R_2(\theta_{13})
P_3(-\delta_\ell) R_3(\theta_{12}) ,
\end{array}
\eqno(4.12)
$$
where the rotation angles are fixed by the observed 
CKM mixing data as
$$
\begin{array}{l}
\theta_{13} =\sin^{-1} |V_{ub}| , \\
\theta_{23} =\sin^{-1} (|V_{cb}|/\sqrt{1-|V_{ub}|^2}) , \\
\theta_{12} =\sin^{-1} (|V_{us}|/\sqrt{1-|V_{ub}|^2}) , 
\end{array}
\eqno(4.13)
$$
and the phase parameters are taken as 
$\delta_q \simeq 70^\circ$ and $\delta_\ell = 180^\circ$.
This suggests that the mass matrices $M_d$ and $M_e$ 
in the $x$-basis  are
diagonalized by the same rotation matrix 
$$
R_d = R_2(\theta_{13}^d) R_3(\theta_{12}) ,
\eqno(4.14)
$$
while the up-quark mass matrix $M_u$ in the $x$-basis is
diagonalized by
$$
R_u = R_2^T(\theta_{13}^u) R_1^T(\theta_{23}) ,
\eqno(4.15)
$$
where $\theta_{13}=\theta_{13}^d -\theta_{13}^u$, 
$\theta_{23}$ and $\theta_{12}$ are given by (4.13),
and the phase parameters are given by 
$$
\delta_q=\delta_d-\delta_u \simeq 70^\circ , \ \ \ 
\delta_\ell=\delta_e-\delta_u = 180^\circ .
\eqno(4.16)
$$
(Since we have assumed that $Y_u$ and
$Y_e$ are real in the present model, the phase factors 
$\delta_u$ and $\delta_e$ must be $0$ or $\pi$.) 
Therefore, forms of the mass matrices $M_u$,
$M_d$ and $M_e$ in the $x$-basis are given by 
$$
\begin{array}{l}
\langle Y_u \rangle_x = P_2(\delta_u) R_2^T(\theta_{13}^u) 
R_1^T(\theta_{23}) D_u R_1^T(\theta_{23}) R_2(\theta_{13}^u)  
P_2(\delta_u) , \\
\langle Y_d \rangle_x = P_2(\delta_d) R_2(\theta_{13}^d) 
R_3(\theta_{12}) D_u R_3^T(\theta_{12}) R_2^T(\theta_{13}^d)  
P_2(\delta_d) , \\
\langle Y_e \rangle_x = P_2(\delta_e) R_2(\theta_{13}^d) 
R_3(\theta_{12}) D_e R_3^T(\theta_{12}) R_2^T(\theta_{13}^d)  
P_2(\delta_e) . 
\end{array}
\eqno(4.17)
$$
In other words, one can choose such a $x$-basis in which 
the mass matrices $M_d$ and $M_e$ are diagonalized
simultaneously, and CP-violating phase factors are 
factorized as shown in (4.17).

\vspace{3mm}

{\large\bf 5 \ Concluding remarks}

When one consider a neutrino mass matrix form
$$
M_\nu \propto \left(\langle \Phi_e \rangle^m \langle \Phi_u \rangle^n
+ \langle \Phi_u \rangle^n \langle \Phi_e \rangle^m \right)^{-1} ,
\eqno(5.1)
$$
one can find that a case which can give a reasonable lepton mixing
is only a case with $m=-2$ and $n=1$, even if one consider any form
of $U_{ue}$.
(This is related to the observed fact 
$\sqrt{m_c/m_t} \simeq m_\mu/m_\tau$.)
One also find that the case with $m=-2$ and $n=1$ can lead to a 
nearly tribimaximal mixing only when one assume 
$U_{ue} =V_{SD}(\pi)$, where $V_{SD}(\delta)$ is the standard 
expression of the CKM matrix with the inputs $|V_{us}|$, 
$|V_{cb}|$ and $|V_{ub}|$.
Therefore, in the present paper, it has been investigated what
structure of the neutrino mass matrix form (1.2) play an
essential role in giving a nearly tribimaximal mixing.
We have found that, in order to obtain such a nearly 
tribimaximal mixing, we need to accept the three 
phenomenological relations (3.9), (3.10) and (3.16).
It is hard to consider that such the relations 
accidentally hold, so that we consider that
the ad hoc assumption $U_{ue} =V(\delta_\ell)$ has
an underlying meaning.
In Sec.4, we have investigated  possible structures of
the down-quark and charged lepton mass matrices.
We have concluded that there must be a specific flavor
basis in which the down-quark and charged mass matrices
are simultaneously diagonalized.

In the present model, an O(3) flavor symmetry has
been assumed.
Relations which are obtained from the O(3)$_F$ invariant 
superpotential by using SUSY vacuum conditions hold only
in flavor bases which are connected by an orthogonal
transformation.
Therefore, in order to use those relations in the
$e$-basis and/or $u$-basis, it has been assumed
that $\langle \Phi_e \rangle$ and $\langle \Phi_u \rangle$
are real and the $e$-basis and $u$-basis can be connected
by an orthogonal transformation $U_{ue}$.
On the other hand, one knows that $\langle Y_d \rangle$
cannot be real because of the observation of $CP$ violating
phenomena in the quark sector.
Therefore, one cannot use the relations from the SUSY 
vacuum conditions in the $d$-basis.
(However, this does not mean that one cannot build
a down-quark mass matrix model.
Relations including Yukawaon $Y_d$ still hold in the
$u$-basis.)

In spite of such disadvantage of the O(3)$_F$ model,
the reason that one consider O(3) flavor symmetry is
as follows:
If we consider a U(3) flavor symmetry, the Yukawaon
$Y_R$ (and also $Y_u$ in a grand unification scenario)
must be {\bf 6} of U(3)$_F$.
It is difficult to build a U(3)$_F$ invariant superpotential
for $Y_R$ without considering higher dimensional terms.
(For example, a Yukawaon model based on a U(3)$_F$ symmetry
is found in Ref.\cite{empiricalMnu08}. However, 
the superpotential term for $Y_R$ in the U(3)$_F$ model
is somewhat intricate.)
In order to build a simpler model for $Y_R$, one will be
obliged to adopt an O(3)$_F$ model.

In the present scenario, it is assumed that there are no 
higher dimensional terms with $(1/\Lambda)^n$ ($n \geq 1$)
in the superpotential except for the effective Yukawa
interaction terms $W_Y$, Eq.(2.1).
Although we want to build a model of $W_Y$ without
any higher dimensional terms, at present, we have
no idea for such a model.
It is a future task to us.

So far, we have not discussed a structure of $W_d$ which
gives a down-quark mass matrix $\langle Y_d\rangle$,
although an attempt to give such $W_d$ has been 
proposed in Ref.\cite{Koide-O3}.
Since this is not the question of the moment in the 
present paper, we did not discuss.
We will discuss a possible structure of $W_d$ elsewhere.

Although the present approach to the masses and mixings 
of quarks and leptons is not conventional and not yet
established, 
this approach will become one of the promising approaches
because  one can treat the masses and mixings without
discussing explicit forms of the Yukawa coupling constants.

\vspace{6mm}

\centerline{\large\bf Acknowledgments} 

The author would like to thank T.~Yamashita for 
helpful discussions, especially on the O(3) symmetry.
This work is supported by the Grant-in-Aid for
Scientific Research, Ministry of Education, Science and 
Culture, Japan (No.18540284).

\vspace{4mm}

\centerline{\large\bf Appendix A: An example of $W_{\Phi f}$} 

The superpotential term $W_{\Phi u}$ in Eq.(2.2) has been 
introduced to fix the VEV spectrum of the ur-Yukawaon $\Phi_u$. 
In this appendix, we demonstrate an example of $W_{\Phi u}$.

When one introduces a further new field $B_u$ with a U(1)$_X$
charge $Q_X= -(3/2) x_u$, one can have a term 
${\rm Tr}[\Phi_u Y_u B_u]$.
However, of course, if one has only this term, one cannot fix
the eigenvalues of $\langle \Phi_u \rangle$, because one needs
a cubic equation in $\langle \Phi_u \rangle$.
Therefore, one assume existence of ${\rm Tr}[A]{\rm Tr}[BC]$, 
${\rm Tr}[B]{\rm Tr}[CA]$ and ${\rm Tr}[C]{\rm Tr}[AB]$ 
in addition to the term ${\rm Tr}[ABC]$ only for the term 
$W_{\Phi u}$. 
Then, the superpotential $W_u$ for the up-quark sector is
given by
$$
W_u = \lambda_u {\rm Tr}[\Phi_u \Phi_u A_{u}]
+\mu_u {\rm Tr}[Y_u A_{u}] + W_{\Phi u} ,
\eqno(A.1)
$$
$$
W_{\Phi u} = y_u  {\rm Tr}[(\Phi_u Y_u +Y_u \Phi_u ) B_{u}]
+ 2y_{1u}  {\rm Tr}[\Phi_u]  {\rm Tr}[Y_u B_{u}]
$$
$$
+  2y_{2u}  {\rm Tr}[Y_u]  {\rm Tr}[\Phi_u B_{u}]
+ 2y_{3u}  {\rm Tr}[B_u]  {\rm Tr}[\Phi_u Y_{u}] .
\eqno(A.2)
$$

The SUSY vacuum condition $\partial W/\partial A_u=0$
has already been investigated in Sec.2.
In this appendix, we will investigate 
$\partial W/\partial Y_u=0$, $\partial W/\partial \Phi_u=0$
and $\partial W/\partial B_u=0$.

From the conditions  $\partial W/\partial Y_u=0$ and 
$\partial W/\partial \Phi_u=0$, one obtains
$$
\frac{\partial W}{\partial Y_u} = 0
= \mu_u A_{u} + y_u (\Phi_u B_u + B_u \Phi_u)
+2y_{1u} {\rm Tr}[\Phi] \, B_u 
+2y_{2u} {\rm Tr}[\Phi B_u ] \, {\bf 1} 
+2y_{3u} {\rm Tr}[B_u] \, \Phi_u ,
\eqno(A.3)
$$
$$
\frac{\partial W}{\partial \Phi_{u}} = 0
= \lambda_u (\Phi_u A_{u} + A_{u} \Phi_{u})
+ y_u (Y_u B_u + B_u Y_u)
+y_{1u} {\rm Tr}[Y_u B_u] \, {\bf 1} 
+y_{2u} {\rm Tr}[Y_u ] \, B_u
+y_{3u} {\rm Tr}[B_u] \, Y_u .
\eqno(A.4)
$$
Since one searches a vacuum with $\Phi_u \neq 0$ and $Y_u \neq 0$,
one can obtain
$$
A_u = B_u =0 ,
\eqno(A.5)
$$
by requiring Eqs.(A.3) and (A.4) simultaneously.
On the other hand, from $\partial W/\partial B_u=0$, 
one obtains
$$
\frac{\partial W}{\partial B_{u}} = 0
= y_u (\Phi_u Y_{u} + Y_{u} \Phi_{u})
+2y_{1u} {\rm Tr}[\Phi_u ] \, Y_u 
+2y_{2u} {\rm Tr}[Y_u ] \, \Phi_u
+2y_{3u} {\rm Tr}[\Phi_u Y_u] \, {\bf 1} .
\eqno(A.6)
$$
By substituting $Y_u \propto \Phi_u \Phi_u$, Eq.(2.4),
one obtains a cubic equation in $\Phi_u$:
$$
 y_u \Phi_u^3 + y_{1u} {\rm Tr}[\Phi_u ]\, \Phi_u^2
+ y_{2u}  {\rm Tr}[\Phi_u^2]\, \Phi_u
+ y_{3u}  {\rm Tr}[\Phi_u^3]\, {\bf 1} =0.
\eqno(A.7)
$$
Since the coefficient of $\Phi_u$, 
$y_{1u} {\rm Tr}[\Phi_u ]/2 y_u $, 
in a cubic equation (A.7) must be equal to $-{\rm Tr}[\Phi_u ]$,
one obtains a restriction
$$
y_{1u}= - y_u.
\eqno(A.8)
$$
Also, from constraints for the coefficients of $\Phi$ and 
${\bf 1}$ in the cubic equation,
one obtains
$$
\frac{y_{2u}}{y_u} {\rm Tr}[\Phi_u^2] = \frac{1}{2}
\left( {\rm Tr}[\Phi_u]^2 - {\rm Tr}[\Phi_u^2] \right) ,
\eqno(A.9)
$$
and
$$
\frac{y_{3u}}{y_u} {\rm Tr}[\Phi_u^3] = -{\rm det}\Phi_u ,
\eqno(A.10)
$$
respectively.
The constraints (A.9) and (A.10) lead to formulas
$$
\frac{ {\rm Tr}[\Phi_u^2] }{{\rm Tr}[\Phi_u]^2}=
\frac{1}{1+2y_{2u}/y_u} ,
\eqno(A.11)
$$
and
$$
{\rm det}\Phi_u =  \frac{y_{3u}/y_u}{2(1+3 y_{3u}/y_u)} 
{\rm Tr}[\Phi_u] \left( {\rm Tr}[\Phi_u]^2- 
3{\rm Tr}[\Phi_u^2] \right)   ,
\eqno(A.12)
$$
respectively.
Thus, the VEV spectrum can completely be determined by the 
coefficients $y_{1u}/y_u$,  $y_{2u}/y_u$ and $y_{3u}/y_u$.

We also assume the same structure $W_e$ as $W_u$ for the 
charged lepton sector.
Then, if one takes $y_{2e}/y_e=1/4$, one obtains
${\rm Tr}[\Phi_e^2] /{\rm Tr}[\Phi_e]^2=2/3$, so that
one can obtain an interesting charged lepton mass relation
\cite{Koidemass}.
However, since the purpose of the present paper is not to 
discuss the mass spectra of quarks and leptons, we do not
touch this problem.

\vspace{4mm}

\centerline{\large\bf Appendix B: 
Mass matrix form for a tribimaximal mixing} 

A general mass matrix form which gives a tribimaximal mixing
\cite{tribi} has been given by He and Zee \cite{HeZee}. 
We summarize the general form for a case of the tribimaximal 
mixing matrix with phases, and we discuss conditions for
$\sin^2 2\theta_{23}=1$ and $\tan^2 \theta_{12}=1/2$ separately.

An orthogonal mixing matrix $U$ which gives a maximal 
$2\leftrightarrow 3$ mixing 
$$
\sin^2 2\theta_{23}=1 \ \ {\rm and} \ \ U_{13}=0 ,
\eqno(B.1)
$$
is given by a form
$$
\widetilde{U} = \left(
\begin{array}{ccc}
c & s & 0 \\
-\frac{1}{\sqrt2} s & \frac{1}{\sqrt2} c & -\frac{1}{\sqrt2} \\
-\frac{1}{\sqrt2} s & \frac{1}{\sqrt2} c & \frac{1}{\sqrt2} \\
\end{array} \right) ,
\eqno(B.2)
$$
where $c=\cos\theta$ and $s=\sin\theta$.
Since a mixing matrix $U$ with $U_{13}=0$ cannot contain 
a $CP$ violating phase, an extended form $U$ from the 
orthogonal mixing matrix $\widetilde{U}$ to a unitary 
mixing matrix is given by
$$
U = P(\alpha) \widetilde{U} P(\beta) ,
\eqno(B.3)
$$
where 
$$
P(\delta) ={\rm diag}(e^{i \delta_1}, e^{i \delta_2},
e^{i \delta_3}) .
\eqno(B.4)
$$
When one defines a mass matrix $M$ with $M^T =M$ which is 
diagonalized by $U$ as follows:
$$
U^T M U = D \equiv {\rm diag}(m_1, m_2, m_3) ,
\eqno(B.5)
$$
one can obtain
$$
\widetilde{U}^T \widetilde{M} \widetilde{U}
= P^2(-\beta)\, D \equiv {\rm diag}(\widetilde{m}_1, 
\widetilde{m}_2, \widetilde{m}_3) \equiv \widetilde{D},
\eqno(B.6)
$$
where
$$
\widetilde{M} = P(\alpha) M P(\alpha) .
\eqno(B.7)
$$
The matrix $\widetilde{M}$ which is diagonalized by 
an orthogonal matrix is real except for a common phase
factor, so that the eigenvalues $\widetilde{m}_i$ are
also real. 
As seen in Eq.(B.6), the phases $\beta_i$ in
$\widetilde{m}_i = m_i e^{-2 i\beta_i}$ are the so-called
Majorana phases, they are unobservable in neutrino
oscillation experiments.
Hereafter, for convenience, we denote 
$\widetilde{m}_i$ as $m_i$ simply.
Then, one can obtain the explicit form of $\widetilde{M}$ from
$\widetilde{M}=\widetilde{U} \widetilde{D}\widetilde{U}^T$ as
$$
\begin{array}{l}
\widetilde{M}_{11} = \frac{1}{2} (m_2+ m_1)
-\frac{1}{2} (m_2 -m_1) \cos 2\theta , \\
\widetilde{M}_{22} = \widetilde{M}_{33} =
\frac{1}{2} m_3 +\frac{1}{4} (m_2+m_1) 
+\frac{1}{4} (m_2-m_1) \cos 2\theta , \\
\widetilde{M}_{12} = \widetilde{M}_{13} =
\frac{1}{2\sqrt2} (m_2-m_1) \sin 2\theta , \\
\widetilde{M}_{23} = -\frac{1}{2} m_3
+\frac{1}{4} (m_2+m_1) +\frac{1}{4} (m_2-m_1)
\cos 2\theta .
\end{array}
\eqno(B.8)
$$
Therefore, the conditions that the mass matrix 
$\widetilde{M}$ gives the maximal $2\leftrightarrow 3$ 
mixing (B.1) are
$$ 
\widetilde{M}_{12} = \widetilde{M}_{13} \ \ {\rm and} \  \ 
\widetilde{M}_{22} = \widetilde{M}_{33} ,
\eqno(B.9)
$$
i.e.
$$ 
{M}_{12} e^{i\alpha_2} = {M}_{13} e^{i\alpha_3} \ \ {\rm and} 
\  \ 
{M}_{22} e^{2i\alpha_2} = {M}_{33} e^{2i\alpha_3}.
\eqno(B.10)
$$
The conditions (B.10) are rewritten as
$$
\left( \frac{M_{12}}{M_{13}} \right)^2 =
\frac{M_{22}}{M_{33}} = e^{2i(\alpha_3-\alpha_2)} .
\eqno(B.11)
$$

On the other hand, the mixing angle $\theta \equiv \theta_{12}$
is obtained from
$$
\tan 2\theta = \frac{2\sqrt2 \widetilde{M}_{12}}{
\widetilde{M}_{33} + \widetilde{M}_{23}  -\widetilde{M}_{11} } ,
\eqno(B.12)
$$
i.e.
$$
\tan 2\theta = \frac{ 2\sqrt2 \eta ({M}_{12} M_{13})^{1/2}}{
\eta^2 \left( (M_{22} M_{33} )^{1/2} +M_{23} \right)
-M_{11} } ,
\eqno(B.13)
$$ 
where
$$
\eta= \exp i\left( -\alpha_1 + \frac{\alpha_2+\alpha_3}{2}  \right) .
\eqno(B.14)
$$
Therefore, the conditions for a tribimaximal mixing, i.e. 
constraints (B.1) and 
$$
\tan^2 \theta = \frac{1}{2} ,
\eqno(B.15)
$$
require the conditions (B.11) and 
$$
\eta^2 \left( (M_{22} M_{33} )^{1/2} +M_{23} \right) -M_{11}
=\eta ({M}_{12} M_{13})^{1/2} ,
\eqno(B.16)
$$
respectively.

\vspace{4mm}


\end{document}